\documentclass[a4paper]{jpconf}
\usepackage{graphicx}
\usepackage{slashed}
\usepackage{amsmath}
\usepackage[usenames,dvipsnames]{color}
\usepackage[colorlinks=true, urlcolor=blue, linkcolor=blue, 
citecolor=blue]{hyperref}

\usepackage{bbm}                                                  
\usepackage{cprotect}

\usepackage{cite}

\usepackage[top=0.8in,bottom=1in,left=0.9in,right=0.9in]{geometry}

\def\ccal{{\cal C}}
\def\fcal{{\cal F}}
\def\gcal{{\cal G}}
\def\lcal{{\cal L}}

\def\ocal{{\cal O}}
\def\scal{{\cal S}}

\def\up#1{^{\left( #1 \right) }}
\def\mati{{\mathbbm1}}

\def\mev{\hbox{MeV}}
\def\gev{\hbox{GeV}}

\def\fm{\fcal}

\def\phitd{\tilde\phi^\dagger}
\def\gdm{\gcal_{\rm DM}}
\def\pmns{V}
\def\msc{m_\Phi}
\def\mfe{m_\Psi}
\def\cw{c_{W}}
\def\sw{s_{W}}
\def\lx{\lambda_x}
\def\gsim{\mathrel{\rlap{\lower4pt\hbox{\hskip1pt$\sim$}}
    \raise1pt\hbox{$>$}}}                
\newcommand{\bea}{\begin{eqnarray}}  \newcommand{\eea}{\end{eqnarray}}
\newcommand{\beq}{\begin{equation}}  \newcommand{\eeq}{\end{equation}}
\newcommand{\bit}{\begin{itemize}}   \newcommand{\eit}{\end{itemize}}

\begin{document}
\title{Dark matter and the neutrino portal paradigm}

\author{V. Gonzalez Mac\'ias$^1$, J. Illana$^2$, J. Wudka$^1$}

\address{$^1$Department of Physics \& Astronomy,  University of California Riverside,  \\ Riverside, CA 92521-0413, USA}
\address{$^2$CAFPE and Depto. de F\'isica Te\'orica y del Cosmos, Universidad de Granada \\
18071 Campus de Fuentenueva, Granada, Spain}

\ead{vanniagm@fisica.ugto.mx,
jillana@ugr.es, 
jose.wudka@ucr.edu}

\begin{abstract}
A simple extension of the Standard Model (SM)  that provides an explicit realization of the dark-matter (DM) neutrino-portal paradigm is presented. The leading interactions between the dark sector, containing scalars and relic fermions, and the SM involve neutrinos. This model meets all observational constraints.
\end{abstract}

\section{Introduction}

Dark Matter (DM) presents an unambiguous evidence of physics beyond the Standard Model (SM). The most compelling DM paradigm assumes that it consists of one or more particles with very weak couplings to the SM \cite{Goodman:1984dc,Feng:2010gw}, and having the correct abundance to explain the CMB observations \cite{Ade:2013zuv}. This hypothesis has been probed extensively using direct \cite{Akerib:2013tjd,Aprile:2012nq,Ahmed:2009zw} and indirect detection \cite{Choi:2015ara,Aartsen:2014hva,Ackermann:2015zua,Aharonian:2006wh,Aguilar:2013qda,Adriani:2008zr} experiments, and in collider processes \cite{Khachatryan:2014rra,ATLAS:2012ky,Goodman:2010ku,Bai:2010hh}. To date, no evidence of DM effects in any of these experiments has been confirmed.

Many viable DM candidates have been proposed, spanning a large range in masses and interaction strengths (see, for example,  \cite{Feng:2010gw,Bertone:2004pz} ). In particular, reference \cite{Macias:2015cna} describes a possible scenario that ensures  naturally small direct and indirect detection signals, without compromising the relic abundance inferred from CMB experiments. This scenario is based on the assumption that interactions between the dark and SM sectors are mediated by one or more Dirac fermions $\fm$, assumed neutral under all dark and SM symmetries, except fermion number. In addition, the dark sector is assumed to contain (at least) one fermion $ \Psi $ and one scalar  $ \Phi $ that have the same (non trivial) transformations under a symmetry group $\gdm $, whose nature is not necessarily specified; the only assumption is that all SM particles are singlets under $ \gdm $, which ensures that the lightest dark-sector particle will be stable and so serve as a DM candidate. 

Reference \cite{Gonzalez-Macias:2016vxy}  shows the simplest model that realizes such a scenario and the implications of existing and projected experimental restrictions on the model parameters. Despite the high precision constraints available, there are significant regions of parameter space allowed. Given that the leading couplings between the DM and the SM sectors involve neutrinos, the model has a distinctive identifying feature: the presence of a monochromatic neutrino signal generated by DM annihilation in astrophysical objects.

\section{Neutrino portal in effective theories}

A dark sector that contains scalars $ \Phi $ and fermions $ \Psi $ allows for the presence of an effective interaction with the SM of the form
 $
\ocal\up5 = (\bar\Psi \Phi)(\phitd \ell) 
\label{eq:d5}
$
where $ \ell $ and $ \phi $ denote, respectively, the isodoublets for a left-handed SM lepton and SM scalar ($\tilde\phi=i\sigma^2\phi^*$); this dimension-5 operator can be generated at tree-level by the exchange of a neutral fermion $ \fm $. Within the $\fcal$-mediated paradigm this operator describes the strongest interactions between the SM and the dark sector, which always involve a neutrino: this is a neutrino portal scenario (neutrino portals have been studied in related contexts for example in \cite{NeuPortal}). The presence of a $\Psi$-$\Phi$-$\nu$ coupling also implies that the heaviest of the dark particles will promptly decay into the lightest, so there will be a single DM relic despite having a dark sector with two (or more) particles. 

In addition to the above $\fm$-induced coupling, the presence of dark scalars allows for the usual Higgs portal coupling $ |\Phi|^2 |\phi|^2 $. If the dark fermion is heavier than the dark scalar, $ \mfe > \msc $, then $\Phi$ constitutes the DM relic and the physics of the model is dominated by the effects of the Higgs portal coupling, which has been extensively studied in the literature. In contrast, if $ \mfe < \msc $, the Higgs portal coupling is secondary to (\ref{eq:d5}) and the phenomenology is completely different; for example, the leading interactions relevant for direct detection are produced by the dimension 5: $|\phi|^2 \bar\Psi \Psi$ and 6: $ (\phi^\dagger\!\!\stackrel\leftrightarrow{D_\mu}\!\!\phi ) \left(\bar\Psi \gamma^\mu P_{L,R}\Psi \right)
\,,\,( \bar\ell \gamma_\mu \ell ) (\bar\Psi \gamma^\mu P_{L,R}\Psi )
$ effective operators that are generated at one loop\footnote{Current-current operators involving quarks or right-handed leptons are only generated at $\ge2$ loops.} by the $ \fm$. The fact that the $ \fm $ create interactions at tree-level and  at one loop is what  allows for the required relic abundance to be obtained within the constraints of direct and indirect detection experiments, without fine-tuning.

\section{UV completion: Neutrino portal DM model} 

The simplest model realization (fully described in \cite{Gonzalez-Macias:2016vxy}) of the above paradigm consists of the following: i) A dark sector which contains one scalar $\Phi$ and one fermion field $\Psi$, transforming under a global symmetry under which all SM fields are singlets such that $ \msc > \mfe $, so that the fermion is stable.  ii) The model contains in addition three Dirac fermions $\fm$, neutral under the dark and the SM symmetries. iii) Lepton number is conserved (except for possible instanton effects). The Lagrangian of the model is:
\bea
\lcal &=& \bar\ell i \slashed{D} \ell + \overline{ e_R } i \slashed D e_R + \bar\Psi(i \slashed\partial - \mfe) \Psi + \bar\fm(i \slashed\partial - M) \fm + |\partial\Phi|^2 - \msc^2 |\Phi|^2 \cr
&& \quad - \left( \bar\ell Y\up e e_R \phi + \bar\ell Y\up\nu \fm \tilde\phi + \bar\Psi  z^\dagger \fm \Phi +  {\rm H.c.} \right)-\lambda_x|\Phi|^2|\phi|^2
\label{eq:lag}
\eea
where $ \ell_i$ and $e_{R\,i}$ denote, respectively, the left-handed SM lepton isodoublets and right-handed isosinglets ($i=1,2,3$ is a family index); $ \phi$ is the SM scalar isodoublet; $ \Psi$ and $ \Phi$ are the dark fermion and scalar fields; and $ \fm_i$ are the (Dirac) neutral fermion mediators.  $M$ is the $ 3 \times 3 $ Hermitian mass matrix for the $\fm$, while the  Yukawa couplings $Y\up e, Y\up\nu$ are general $3\times 3$ complex matrices.

The fields $ \fm$ and $ \nu $ are replaced by the fields $N$ and $n_L$ to diagonalize the mass matrix:
\beq
\fcal = U^\dagger (\ccal U_L N_L -  \scal n_L + U_R N_R )\,,\qquad
\nu = \pmns(\scal U_L N_L + \ccal n_L)
\label{eq:fieldrot}
\eeq
where 
$ Y\up\nu =\sqrt{2}/v \pmns \eta U M$ is redefined with a standard polar decomposition, and $\ccal =1/(\mati + \eta^2 )^{1/2}$, $\scal = \eta/(\mati + \eta^2)^{1/2}$. The unitary matrices $U_{L,R} $ are chosen such that
$
U_R^\dagger  U M U^\dagger \ccal^{-1} U_L = M_N = {\rm diagonal}\,.
$
In this basis the $n_L$ are massless left-handed fermions that correspond to the SM neutrinos, and $(N_L,N_R)$ form a set of Dirac fermions with mass matrix $M_N$. The interaction terms then become:
\bea
&-&\lcal_{\rm int} =   \left(H/v \right) \bar e m_e e + \lx |\phi|^2 |\Phi|^2 + \frac g{2\cw} \bar e \slashed Z ( 2 \sw^2 - P_L) e \cr
&+& \left[\bar\Psi  z^T U^\dagger( \ccal U_L N_L - \scal n_L + U_R N_R)  \Phi +  {\rm H.c.} \right] + \frac g{\sqrt{2}} \left[ \bar e \slashed W \pmns  (\ccal n_L + \scal U_L N_L) + {\rm H.c.} \right] \cr
 &+& \left(H/v \right) \left[  \bar N_R M_N U_L^\dagger \scal (\ccal n_L + \scal U_L N_L )   +  {\rm H.c.} \right]  + \frac g{2\cw}  (\bar n_L \ccal +  \bar N_L U_L^\dagger \scal) \slashed Z (\ccal n_L + \scal U_L N_L)\,
\label{intlag}
\eea
The number of physical parameters is 3 for $ m_e,\,M,\,z$ and $ \eta$ each, plus 1 for $\mfe,\,\msc,\,\lx $ each, plus 4 for $\pmns$ and 8 for $U$, for a total of 27. 

Note that the presence of the $ \fcal $ in (\ref{intlag}) modifies the couplings of the charged and light neutral leptons to the SM bosons, as well as couplings involving the heavy neutral leptons $N$. The DM-SM interactions arise from the fourth term in (\ref{intlag}).

\begin{figure}[th]
\centerline{
\includegraphics{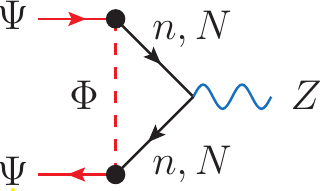}\qquad\,\, \,\includegraphics{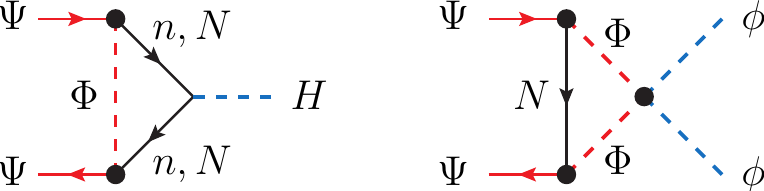}
\label{fig:PP1L}
}
\caption{Interaction of the $Z$ (left) and $H$ bosons (center \& right)  with the DM fields induced at one loop.\label{fig:PPZ1L}}
\end{figure}

As emphasized above, the term $\bar\Psi  z^T U^\dagger  \scal n_L   \Phi$, represents the leading couplings between the dark and SM sectors, which justifies our denoting this a ``neutrino portal'' scenario. The presence of the $ \Psi$-$\Phi$-$n_L$ coupling implies that whenever $ \msc > \mfe $ the scalar field will decay promptly into the fermion and a neutrino:\footnote{If $ \mfe > \msc$ it is the fermions that decay.} this model, while having a multi-component dark sector, has a single component DM relic. However, the presence of the $ \Phi $ is essential for generating the leading DM-SM interactions. Although $Z$-DM and $H$-DM couplings are not generated at tree level in (\ref{intlag}), they are induced at one loop (fig. \ref{fig:PP1L}) and represent the leading coupling in direct detection processes and produce important resonant effects in the annihilation cross section when $m_\Psi\simeq m_Z/2$.

In the special case, thoroughly revised in \cite{Gonzalez-Macias:2016vxy} , where $ |\eta|\ll1 $ and the mass states $N$ are almost degenerate:  $M \simeq \Lambda [\mati+\ocal (\eta^2)] $. This implies that, up to $\ocal (\eta^2)$ corrections, $ U_{L} = U_{R}=U $, $\ccal \simeq \mati - \eta^2/2 $, $\scal \simeq \eta $ and $ M \simeq \Lambda(\mati-\frac{1}{2}U^\dagger\eta^2 U)$.
As a consequence, the observables of interest (the cross sections relevant for relic abundance and indirect detection calculations) will  depend on $\lx,\, \eta,\,U$ and $z$ only through the two real combinations $|\eta U z|$ and $\lx |z|^2$. 

\section{Observational constraints}
All the experimental constraints that are presented below are computed assuming the special case of quasi-degenerate heavy fermions. In this case the relevant parameters are the masses $ \mfe,\, \msc$, a heavy mass scale $ \Lambda $ (the cut-off scale) and the coupling combinations $|\eta U z|$ and $\lx |z|^2$. Two possibilities are analyzed for the mass spectrum in the dark sector,  a quasi-degenerate spectrum $m_\Phi<m_\Psi+10\,\gev$ and a non-degenerate spectrum $m_\Phi\ge\mfe+10\,\gev$, with $ \msc > \mfe $ in either case, as required in the scenario considered here. Additionally, it is required that $|z|\le2$, which is slightly more conservative than the limits $|z_i| < \sqrt{4\pi}$, derived from tree-level unitarity\footnote{We impose tree-level unitarity given our requirement that the model remains perturbative.} for each component of $z$, using the process $ \Psi \fm \to \Psi \fm $. We used the aid of the public codes \verb|MicrOmegas| \cite{Belanger:2013oya} and \verb|CALCHEP| \cite{Belyaev:2012qa,Alloul:2013bka}. 
\subsection{Electroweak constraints}
The tightest restrictions on the model parameters are derived from the decays of the $Z$ and $H$ gauge bosons \footnote{$W$ mediated decays of fermions or neutrino mixings are also modified but are weaker than those from the $Z$ boson decay.}. 
\begin{enumerate}
\item The experimental result $\Gamma(Z\to {\rm inv})=499.0 \pm 1.5\, \mev$ \cite{Agashe:2014kda} for the invisible width of the $Z$ generates the most stringent bound on the parameters of the model:
$\sum_i  \eta_i^2 < 0.014 \quad (3\sigma)$
where the $ \eta_i $ are the diagonal elements of the diagonal matrix $ \eta $. 
\item Latest results from the ATLAS experiment at the LHC \cite{Aad:2015pla,Aad:2015txa,CMS:2015naa} report an upper bound $ \Gamma(H \to {\rm inv}) < 2.2\, \mev $ at a $90\%$ C.L. so that, for $ m_H \gg \mfe $,
\beq
\frac v\Lambda
\left| |\eta U z|^2\frac{\Lambda^2}{v^2}  + \lx |z|^2 \ln\frac\Lambda\msc\right|< 1.7\,.
\label{eq:H_limits}
\eeq
\end{enumerate}
\subsection{Relic abundance}
\begin{figure}[th]
\centerline{
\includegraphics{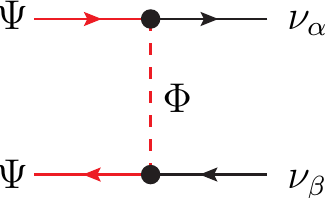}\qquad
\includegraphics{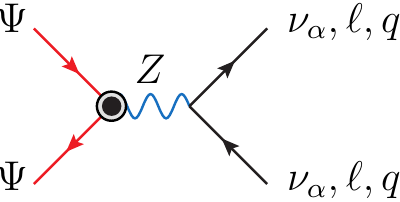}\qquad
\includegraphics{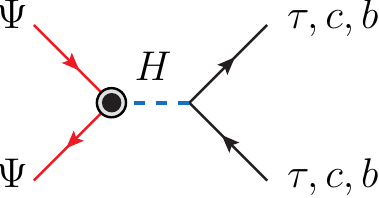}
}
\caption{Leading DM-SM interactions in the annihilation channels. 
\label{fig:PPnn}}
\end{figure}

\begin{figure}[th]
$$\includegraphics[width=.4\textwidth]{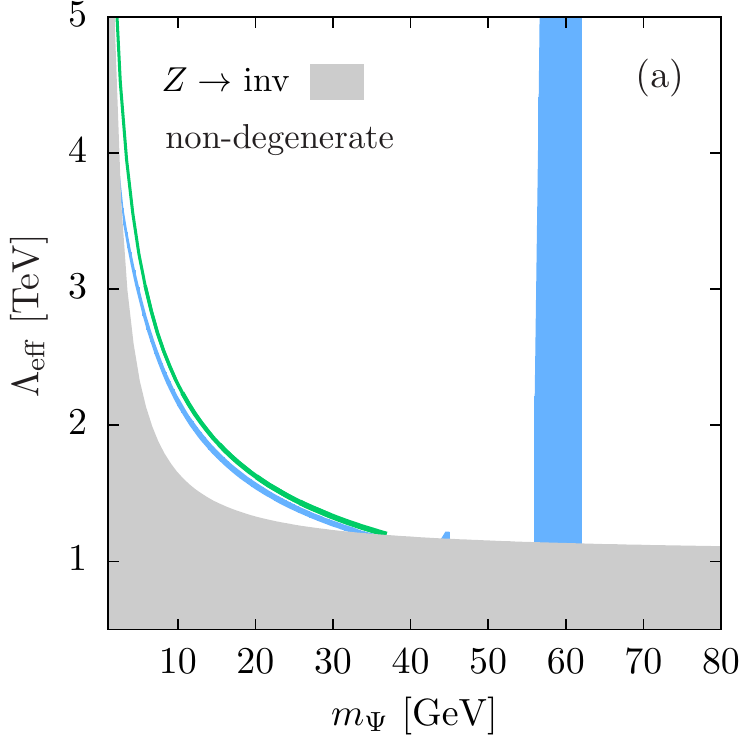}\qquad\includegraphics[width=.4\textwidth]{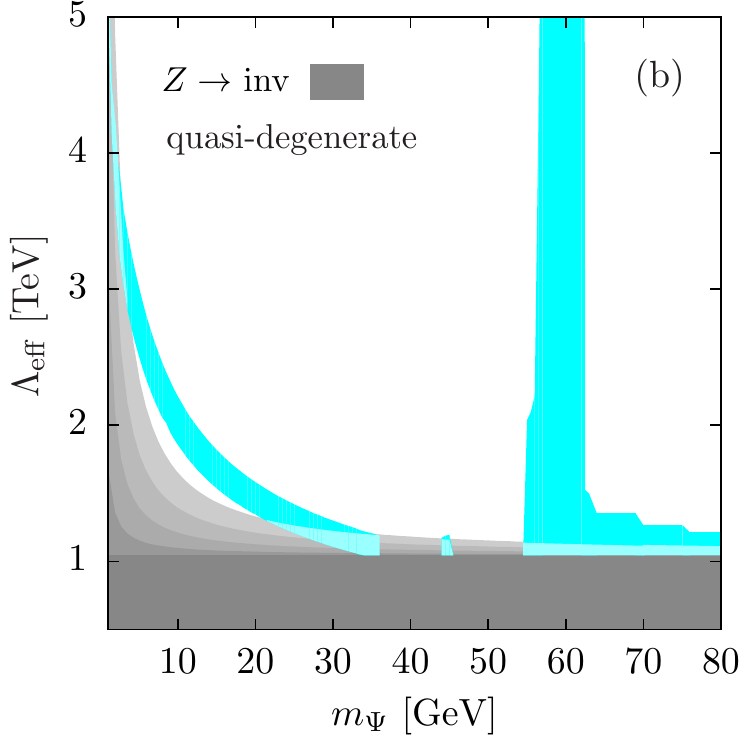} $$
\caption{Constraints on the DM-$\nu$ portal model from the relic abundance $3\sigma$ bounds obtained by the Planck experiment: (a) Allowed regions for the non-degenerate case (blue); for comparison, the green band results from the analytic approximation valid outside the resonance regions. (b) The cyan areas denote the allowed regions for the quasi-degenerate case; in this case the invisible $Z$ constraint is sensitive to $ \msc$ with the grey bands corresponding  to $\msc =\mfe +\{0,1,3,6,10\}\, \mathrm{GeV}$ (dark to light grey, respectively). We set  $|z|=2$ for illustration.}
\label{fig:relic_f}
\end{figure}

The leading DM-SM interaction is generated by the (tree-level) exchange of the dark scalars $ \Phi $ (figure \ref{fig:PPnn}) and represent the most important reaction responsible for the equilibration between the dark and standard sectors in the early Universe. If the dark scalar is slightly heavier than the dark fermion (the so-called the quasi-degenerate case), even though it decays promptly to the dark fermion and a neutrino, coannihilation processes become important when computing the present density of the relic fermions since the temperature in this case at the equilibration process is higher than the difference in their masses. All of these effects are taken into account in the numerical calculations.\footnote{Coannihilation channels for the equilibration process, such as $\Psi\Phi\to We,\,Z\nu,\,H\nu$, and $\Phi\Phi \to HH,WW,ZZ$ become important when kinematically allowed; all are included in the numerical calculations below.} 
The remaining relevant interaction is generated by the one-loop induced $\Psi$-$\Psi$-$H$ coupling, and consists of the $s$-channel exchange of the $H$ boson (figure~\ref{fig:PPnn}), dominant in the resonance region  $ \mfe \simeq m_H/2 $. A similar interaction generated by the $s$-channel exchange of the $Z$ boson is small but observable in the resonance region $ \mfe \simeq m_Z/2 $. The constraints imposed by Planck  \cite{Ade:2013zuv} are better illustrated in the $(\mfe, \Lambda_{\rm eff})$ plane (fig.~\ref{fig:relic_f}), where $\Lambda_{\rm eff}$ defined as $
 \Lambda_{\rm eff} = 
\frac{v/\sqrt{2}}{|\eta U z|}  (\frac{\msc^2+\mfe^2}{\mfe^2})^{1/2}$.

\subsection{Direct detection}
\begin{figure}[th]
\centerline{\includegraphics[width=.6\textwidth]{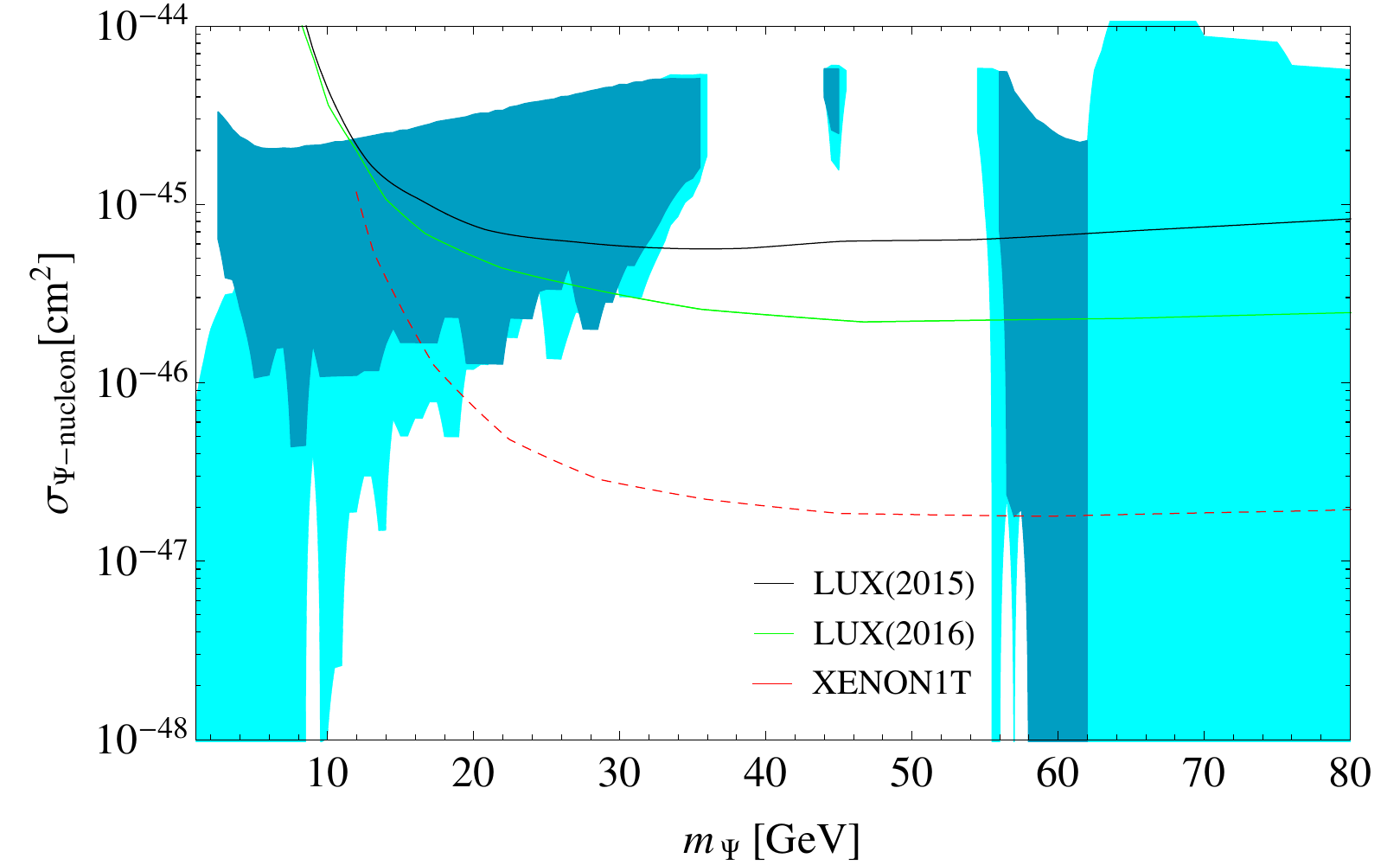}}
\caption{DM-nucleon spin-independent cross sections compatible with the relic abundance and electroweak constraints for the non-degenerate (blue) and quasi-degenerate (cyan) cases. The region above the solid (dashed) lines is (will be) excluded by the LUX (2016) (XENON1T) experiments.
\label{fig:DD}}
\end{figure}

The leading interactions between the dark matter and the neutral bosons $Z$ and $H$ are induced at one loop, generated by the diagrams in figure~\ref{fig:PP1L}  hence, naturally suppressing the interactions of the dark matter with quarks. Both axial-vector and vector and scalar couplings are proportional to $|\eta U z|^2$  or $\lx |z|^2\ln(\Lambda/{m_\Phi})$, where only the first parameter combination is affected by the relic abundance constraints. Hence both spin-dependent and spin-independent cross sections are roughly of the same order. Figure~\ref{fig:DD} shows the projection of the numerical results for xenon nuclei to the $(\mfe, \sigma_{\rm SI})$ plane, together with the present bounds from LUX  \cite{Akerib:2013tjd}\footnote{We present here updated results using the latest LUX bounds (2016) from new analyzed data.} and the expected sensitivity from XENON1T \cite{Aprile:2015zx}.

\subsection{Indirect detection}
\begin{figure}[th]
\centerline{\includegraphics[width=.6\textwidth]{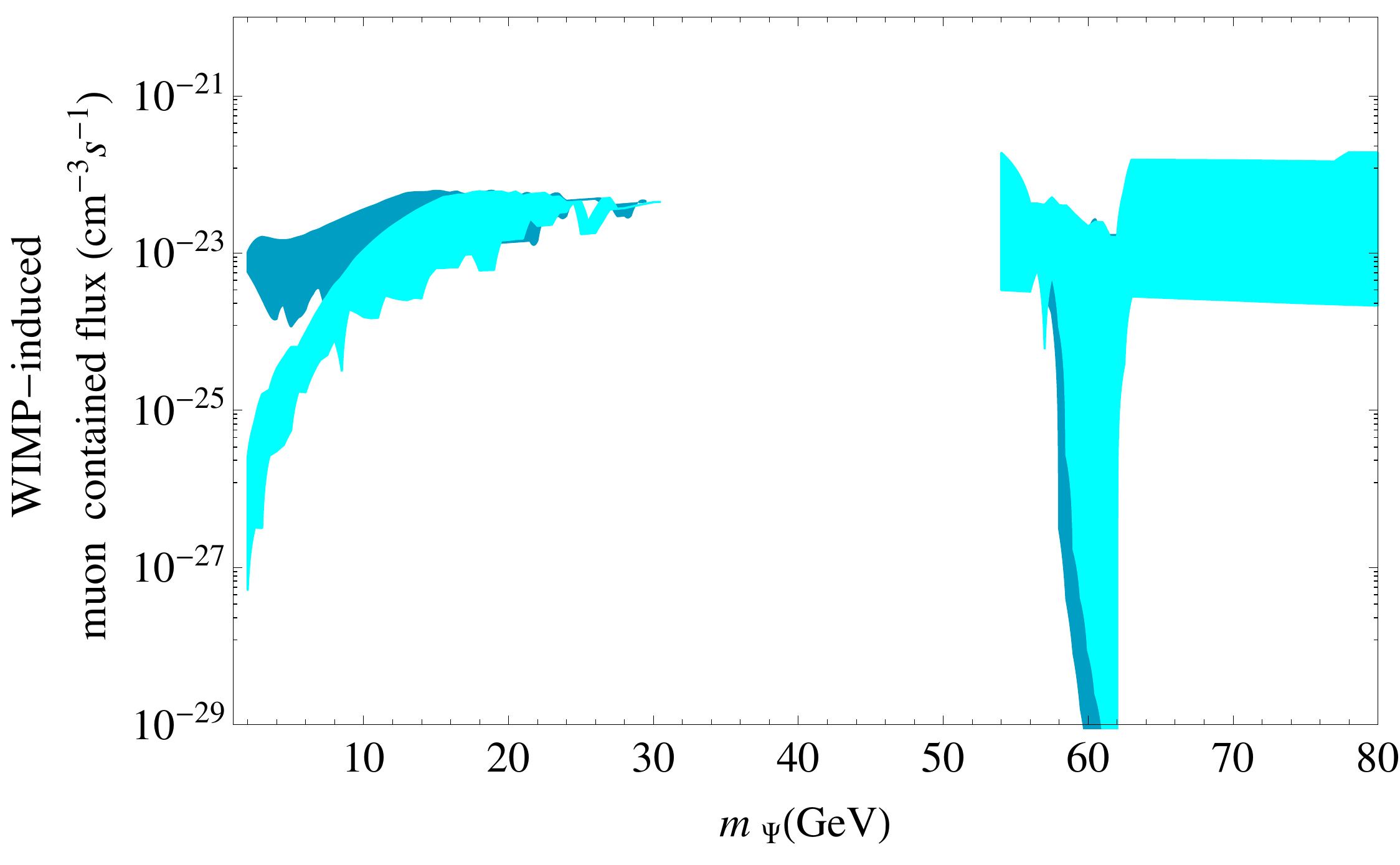}\quad\includegraphics[width=.6\textwidth]{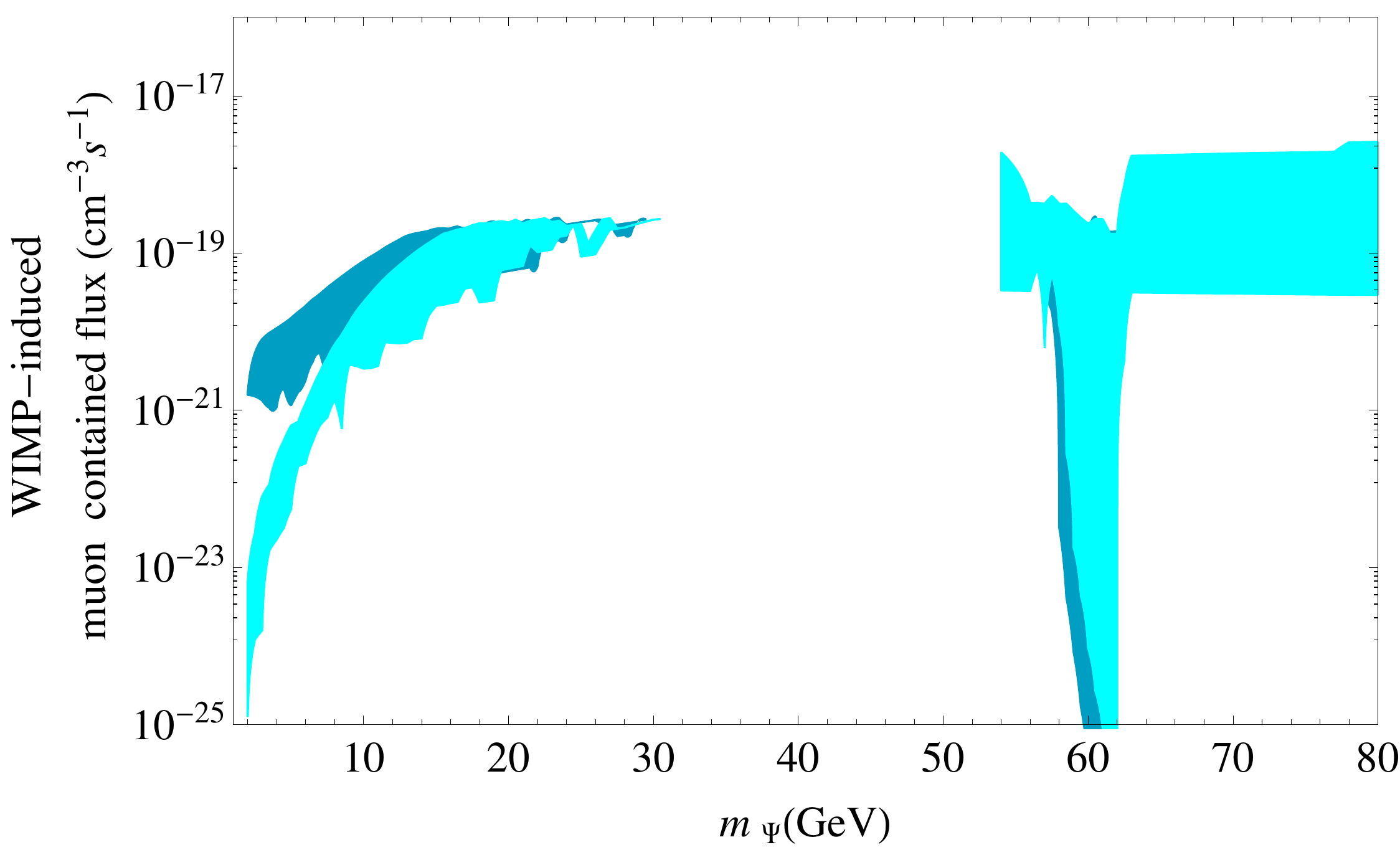}}
\caption{Induced muon rate by neutrinos produced from DM annihilation in the core of the Sun for the non-degenerate (blue) and quasi-degenerate (cyan) cases. The left figure shows the muons produced by neutrinos interacting within the detector (contained), and the right figure shows the induced muon flux by neutrinos interacting with the surrounding material (upward). These figures show the allowed region of parameter space from Planck, Lux (2016), and electroweak constraints.
\label{fig:muonflux}}
\end{figure}

Accumulated DM particles in the core of astronomical objects like the Sun can then annihilate into neutrinos, or other SM particles, that can be detected, after traveling to Earth and interacting with the detectors, in astrophysical high energy neutrino experiments (see for example \cite{Cirelli:2005gh}). Given the small DM velocities, the neutrino spectrum of the $\Psi\bar\Psi\to\nu\bar\nu$ channel in our model is essentially a delta function centered around $E_\nu\simeq \mfe$ that would show up as a monochromatic line in a detector if the corresponding experimental sensitivity is reached. In figure \ref{fig:muonflux} we show the predicted flux of muons produced from the interaction of the neutrinos inside a water Cherenkov detector (contained flux), and the flux of muons produced from the interaction of up-going neutrinos with the rocks surrounding the detector (upward flux), as predicted by our model and taking into account Planck and LUX constraints. There are no significant experimental constraints for neutrino final states in DM annihilation for DM masses below 100~GeV\footnote{Available limits by SuperKamiokande or Icecube \cite{Aartsen:2012kia,Choi:2015ara} depend on the DM annihilation final states, which are chosen usually to be the so-called soft ($b\bar b$) or hard channels ($\tau^+\tau^-$); also available are limits for the $W^+W^-$ and direct neutrino production channels \cite{Belanger:2015hra}}. \\

\begin{figure}[th]
\centerline{\includegraphics[width=.6\textwidth]{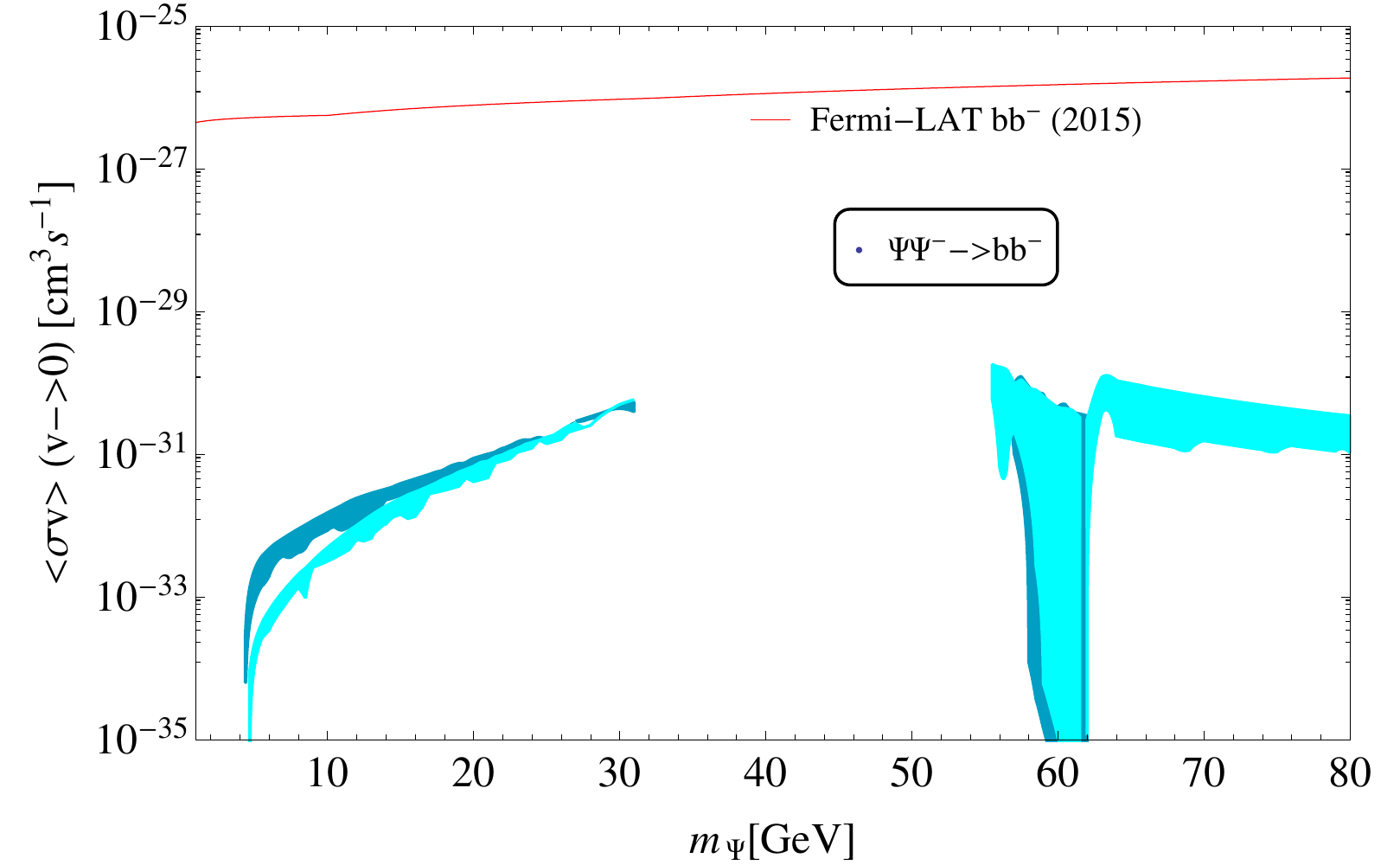}\qquad\includegraphics[width=.6\textwidth]{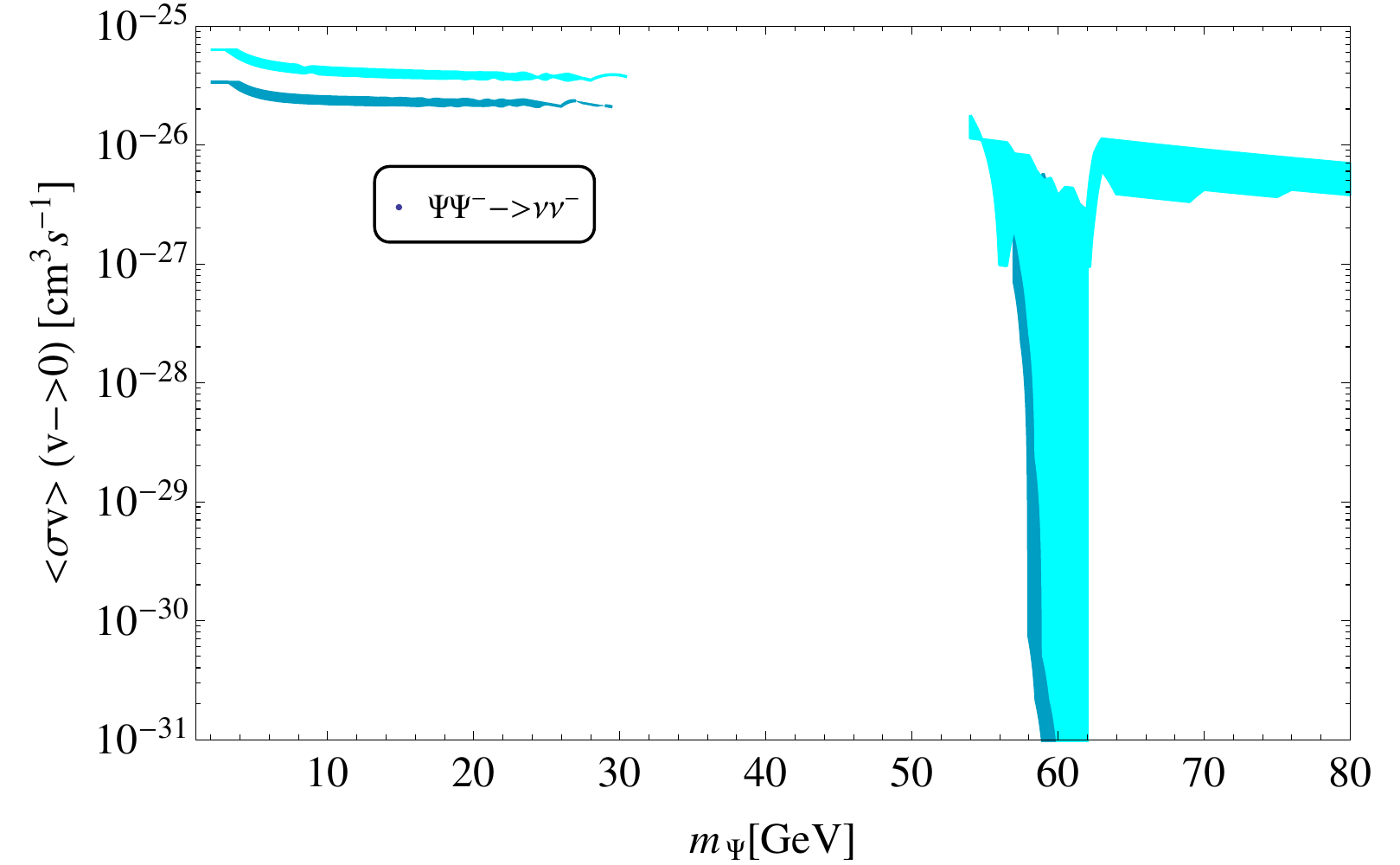}}
\caption{Annihilation cross section into $b$ quarks (left) and $\nu$ (right) final states for the non-degenerate (blue) and quasi-degenerate (cyan) cases. These figures show the allowed region of parameter space from Planck, Lux (2016), and electroweak constraints at $3\sigma$, together with the latest Fermi-LAT results. 
\label{fig:ID}}
\end{figure}

Other extragalactic unresolved point-sources, such as the galactic halo, galactic center, galaxy clusters, dwarf galaxy satellites are also sources of DM annihilation products that may be accessible to indirect detection experiments \cite{Yuksel:2007ac,Bertone:2004pz}. In particular, gamma rays and neutrinos produced as primary or secondary products of DM annihilation will travel essentially undisturbed through space, so the flux of these particles is proportional to the (present time) thermally-averaged, annihilation cross section of  non-relativistic DM relics. In figure \ref{fig:ID} (right) we show the annihilation cross section of $\Psi\bar\Psi$ into neutrinos versus the DM mass $\mfe$ for regions in parameter space that meet the relic abundance and direct detection constraints.
There are no significant experimental constraints for neutrino final states in DM annihilation for DM masses below 100~GeV. Figure \ref{fig:ID} (left) shows the annihilation cross sections of the process $\Psi\bar\Psi\to b\bar b$ in the non-relativistic limit generated by the $Z$ boson exchange, versus the DM mass $\mfe$, with all points fulfilling the relic abundance and direct detection constraints. The recent Fermi-LAT limit~\cite{Ackermann:2015zua}, obtained by searching for $b\bar b$ annihilation products in several dwarf galaxies having a high ratio of DM to ordinary matter, is also displayed.

\section{Conclusions}
A simple model realization for DM interactions in the neutrino portal paradigm is revised taking into account updated LUX results \cite{Akerib:2013tjd}. The relatively large DM-neutrino couplings allow an annihilation cross section large enough to generate the expected relic density, while simultaneously obeying the direct-detection constraints, because of the suppressed couplings to the $Z$ and $H$ without fine tunning. The indirect detection constraints are also easily accommodated because in this scenario the main annihilation products are neutrinos, for which the available limits are weak. It is of interest that there are two distinctive scenarios depending on the mass spectrum in the dark sector. If the dark scalars are only sightly heavier than the fermions, coannihilation processes become important in generating the freeze-out of the DM fermions, the lightest and only stable particles, and wide regions of parameter space are allowed. In the case of non-degenerate dark particle states, electroweak constraints restrict the DM mass to lie in the range  $ 2.3\,\gev\le\mfe \le 35\, \gev $ or near the $H$  resonance region $\mfe\simeq m_H/2$. In contrast, for quasi-degenerate dark scalars and fermions, the electroweak constraints together with LUX constraints exclude only the relatively narrow range $35\,\gev\le\mfe\le 52\,\gev$.

\end{document}